\documentclass[fleqn,10pt]{wlscirep}
\usepackage[utf8]{inputenc}
\usepackage[T1]{fontenc}
\usepackage{nicefrac}
\usepackage{hyperref}

\def\eg{{\em e.g.~}}
\def\CO2{{$\text{CO}_2$\,}}
\def\H2O{{$\text{H}_2\text{O}$\,}}
\newcommand*\norm[1]{\left\lVert#1\right\rVert}

\title{Fire Behavior Monitoring using MeteoSat Third Generation, FCI-FireDyn algorithm: Rate Of Spread and Burnt Area Dynamics for large fire event}

\author[1,*]{Ronan Paugam}
\author[2]{Akli Benali}
\author[3]{Julia Harvie}
\author[4]{Andrea Meraner}
\author[5]{Niels Andela}
\author[6]{Weidong Xu}
\affil[1]{CERTEC, Universitat Politècnica de Catalunya, Barcelona, E08019, Spain, }
\affil[2]{Centro de Estudos Florestais e Laboratório Associado TERRA, Instituto Superior de Agronomia, Universidade de Lisboa, Tapada da Ajuda, 1349-017 Lisboa, Portugal}
\affil[3]{}
\affil[4]{EUMETSAT, Eumetsat Allee 1D-64295 Darmstadt Germany}
\affil[5]{}
\affil[6]{King's College London, Leverhulme Centre for Wildfires, Environment and Society, Department of Geography, Aldwych, London, WC2B 4BG, UK}

\affil[*]{ronan.paugam@upc.edu}

\begin{abstract} 
This study presents FCI-FireDyn, a new algorithm developed to monitor wildfire dynamics using the Flexible Combined Imager (FCI) onboard the Meteosat Third Generation satellite. Leveraging the high temporal resolution of FCI (10-minute full-disk observations), the algorithm derives fire arrival time maps, rate of spread (ROS), and Burn Area (BA) evolution at sub-kilometer spatial resolution and 2-minute temporal intervals. The method combines threshold-based MWIR detection, spatio-temporal interpolation to reconstruct fire front progression and ROS fields at $175$~m resolution. FCI-FireDyn was tested on three major fire events in Southern Europe (Portugal, Greece, and France) from the 2024–2025 seasons. The retrieved BA and Fire Growth Rate show good agreement with reference datasets from EFFIS, Copernicus EMS, and PT-FireSprd, with total final BA deviations below 20\%. The algorithm captures distinct propagation phases, including acceleration episodes that precede FRP peaks, highlighting a potential for NRT fire behavior monitoring. Despite limitations due to FCI spatial resolution, results demonstrate that it provides sufficient spatio-temporal coverage to estimate front-scale fire dynamics. FCI-FireDyn thus represents a proof of concept for deriving high-frequency fire behavior metrics from geostationary observations to support operational and modeling applications.
\end{abstract}
\begin{document}

\flushbottom
\maketitle
%
%
\thispagestyle{empty}

\section*{Introduction}  

Wildfires in fire-prone regions such as western North America, eastern Australia, and the Mediterranean basin have increased over recent decades, leading to significant environmental and economic impacts \cite{Andela2017,Abram2021}. Remote sensing techniques for fire monitoring have been developed for more than two decades, supporting applications in operational fire management and atmospheric modeling.
Satellite platforms have become particularly attractive in this context due to their continuous data availability and their ability to provide observations for very large fire where airborne monitoring can become a challenge.
Fire monitoring at satellite scale is essentially based on 2 products, the Fire Radiative Power (FRP, kW) \cite{Wooster2021} and the Burnt Area (BA, ha) \cite{chuvieco2019}.

From those two products only FRP is capable of providing timely information in Near Real Time (NRT), which usually means in the operational weather forecast community that the product is available within 3 hours after observation. Information on the fire activity (its radiative emitted power in kW) can be converted to gas emission \cite{Wooster2005} which makes FRP very interesting in the context of weather forecast which aims at simulating effects of fire emission on the atmospheric composition. FRP in this case is a gridded product. There are several FRP based fire emission inventory like for example the Global Fire Assimilation System (GFAS) available on a daily basis globally at 0.1 degree resolution \cite{Kaiser2012}.

In the context of fire suppression operations, agencies require information on fire behavior as fast as possible. Rapid access to such data is essential for planning resource allocation and responding to fire evolution.
Fire behaviour refers to the manner in which fire ignites, spreads, and develops over time under varying environmental conditions. Several metrics are used both in operational and research to quantify fire behaviour like the Rate of Spread (ROS, m/s) of the fire front, or the Fire line Intensity (FI, kW/m) \cite{Rothermel1980}. In fire attack, emission are not a priority, however knowing the location of the fire front and its ROS are crucial information to support operitional decision \cite{Monedero2019}. Infrared Images form airborne sensors are a typical source of information for such application \cite{Allison2016}. When not available, satellite data are a good backup and hot spot product such as the one provided by FIRMS are commonly used \cite{DaviesFIRMS2009}. They indicate the presence of a fire at a location set by the center of the satellite image pixel where the fire is detected. In most hotspot product, the FRP is also provided. This time it is now a vector point measurement, and the FRP value can be used to estimate a relative intensity that can give information on the development of the fire.

FRP is a satellite product that has many application in fire monitoring and is listed in the Essential Climate Variable \cite{GCOS2016}. However, when interested in fire behaviour, it is important to note that FRP does not provide information on fire spread. FI, measured in kW/m, is the metric used for fire spread intensity, and FRP and FI are linked via the ROS by the Byram's equation \cite{Johnston2017}.
ROS is therefore a complementary metrics to FRP, that could help operational agency to better characterize fire behavior during fire attack. In the context of emission and analysis of fire behavior at global scale, ROS could also help at better characterizing change in fire regime in ecosystem that are affected by climate change \cite{Cardil2023}.

As FRP, ROS can be estimated using remote sensing \cite{Johnston2018}. It is generally based on airborne infrared data with high temporal resolution and was computed at various scales, ranging from  experiemtal fire (~100 m)\cite{Paugam2013} to landscape scale (1000km) \cite{Stow2019} using infra red sensor in the Middle Wave Infra Red (MWIR) or the Long Wave Infra Red (LWIR) spectral range.

ROS computation is generally divided in a two-steps process: first the fire perimeter needs to be delineated, this provides information of arrival time of the front at a given location. Maps compiling this information are usually referred as arrival time map. A second step use this map to compute ROS at pixel level. A classical approach is for each point on a front, to find its location on the following front, estimate distance and direction of propagation, and compute ROS vector form the time difference between the front \cite{Paugam2013},\cite{Stow2019}. 

ROS computation out of satellite observation is an ongoing action in the fire science community. Thanks to consistent and repeated observations across the entire globe, such product could have application in research work on fire regime \cite{chen2022} or into operational planning as a backup to airborne deployment. Despite a long history of development of Satellite-based active fire monitoring product using MWIR and LWIR infrared observations of fire activity, to our knowledge, satellite ROS product are still not available. The combined temporal and spatial resolution of current polar orbital satellite (\eg VIISRS, Sentinel-3: 300m to 1km, every ~$12~h$) or geostationary satellite (\eg GOES, MSG: several km, repeated time cycle <15min) make it very challenging to compute ROS. 
Although combining overpasses from polar orbital and geostationary satellites is a very attractive approach, to date, there are no reliable harmonized products \cite{Humber2022}, making this easily possible.
The temporal gap present in polar orbital products, where perimeters are available at intervals of $12~h$ can be greatly reduced by combining the multiple satellite now available: the VIIRS sensor is on Suomi, NOAA-20 and NOAA-21 and SLTR is operated on Sentinel-3 A and B. However, gaps still exists in the morning and in the afternoon  after 16h local time \cite{Johnston2020}. On the opposite, geostationary satellite are good at picking the diurnal cycle \cite{Chatzopoulos-Vouzoglanis2023}, however their low spatial resolution is making them less reliable in situations of complex terrain like around waterbodies and mountainous terrain.

Although no operational Rate of Spread (ROS) products are currently available, several studies have analyzed in reconstruction mode historical fire events to characterize fire progression (e.g., sub-daily fire perimeters) \cite{chen2022, Liu2024, Farguell2021a} or to estimate dynamical fire front behavior.
ROS can be quantified at two different spatial scales: at fire-front scale, providing multiple local ROS values along the active perimeter, or at fire scale, describing the overall fire expansion rate.
In fire management applications, at fire scale, the front behavior is often expressed as either the Forward Rate of Spread (FROS) (m/h) or the Fire Growth Rate (FGR) (ha/h) \cite{Benali2023,Duane2024}. FROS represents the temporal average of the fastest propagation along the leading edge of the fire front, while FGR corresponds to the time derivative of the burned area (BA) time series.
In research application and broader scale fire regime analyses, several fire-scale metrics have been proposed to describe overall fire spread behavior, such as the Fire Spread Rate (FSR) [km/day]—defined as the FGR normalized by the fire perimeter \cite{Artes2019,Andela2019}—and the Coarse Rate of Spread (CROS) [km/day], which is based on the most likely propagation path inferred from daily kilometer-scale active fire detections \cite{Humber2022}.

This study aims to present a new algorithm, FCI-FireDyn that aims at estimating fire progression and front scale ROS from observation of the MeteoSat Third Genereation (MTG) geostationary satellite operated by EUMETSAT that were made available in November 2024. 
The core objective of this study is to exploit the high temporal resolution of the Flexible Combined Imager (FCI) onboard the first MTG platform (MTG-I) to derive detailed fire dynamics. Specifically, the approach involves generating arrival time maps from active fire detections, estimating ROS, and subsequently downscaling these arrival time maps through interpolation methods that incorporate an inferred fire propagation direction. The final outcome is a high-resolution (approximately 200~m) and high-frequency (every 2~minutes) characterization of fire spread and burned area evolution.
The FCI sensor onboard MTG-I provides full disc observation of the earth every 10 min at a resolution around 1.5 km at the latitudes of Southern Europe. The data collected from MTG-I are therefore comparable to MODIS observation but delivered every 10 min, making 72 times more observations than with the two MODIS sensors on the same lapse time.
Our objective is to show that the FCI provide a combined spatial and temporal resolution sufficient for this application and that fire behavior metrics such as FGR, ROS and BA can be computed automatically and are comparable with semi-automatic or fully manual approach on specific fires. This work is a proof of concept. It focus on 3 fires scenarios from the 2024 and 2025 fire seasons located in Portugal, Greece and France.

The Results section presents, in reconstruction mode, the performance of the algorithm in computing fire behavior descriptors for the three selected case scenarios. This section primarily provides a qualitative interpretation of the algorithm's outputs. A more detailed quantitative analysis is conducted in the Discussion section, focusing on one of the three fire events for which the data availability supports a time-resolved assessment. The Discussion section also examines the algorithm’s potential, its limitations, and prospects for future development. Finally, the last section provides detailed information on the applied methodology.

\section*{Results}
This section shows results form the FCI-FIreDyn algorithm for three cases study located in Southern Europe: a complex multiple fires merging scenario that occurred between the 15th and the 18th of September 2024 in Portugal near Aveiro, an intense wind driven case that occurred near Varnavas in the Attica region of Greece between the 11th and the 12th of August 2024, and the large fire that occurred in Les Corbi\`{e}res in France near Ribaute between the 5th and 7th of August 2025. 

To enable detailed analysis at the fire front scale, an online visualization platform was developed and is presented in the additional material section. 

\subsection*{Aveiro Fire}
The fire that occurred near Aveiro in September 2024 started on the 15th with two different ignitions that stay apart the first day. A violent outburst happened in the southern fire in the early morning of the 16th ($06:00$ UTC). The fire propagated over the town of Albergaria-a-Velha and later merge with the northern perimeter before spreading further south on the 17th. It burnt until the 18th totaling a surface between $21,262$~ha \cite{EMSR760} and $23,318$~ha \cite{EFFIS2007}, depending of the source, in nearly 4 days. 
Using only images from MWIR band collected by FCI, FCI-FireDyn computes fire progression at a resolution of ($175$~m) and output ROS vector maps, timely BA and associate FGR at a time resolution of $dt=2$~min. The downscaling (increase in both spatial and time resolution) of the fire behavior descriptors is based on fire paths estimated from arrival time map extracted from the original MWIR observation resampled at $768$~m, (see Methods section).
\begin{figure}[!ht]
\centering
\includegraphics[width=\linewidth]{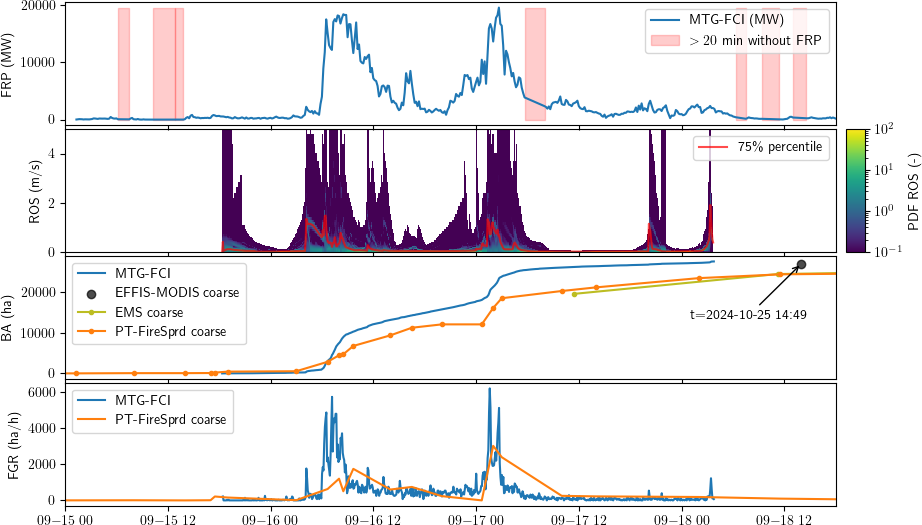}
\caption{Time series of fire-scale fire behavior metrics for the fire that occurred near Aveiro in Portugal between the $15^{\text{th}}$ and the $18^{\text{th}}$ of August 2024. From top to bottom panels: the FRP time series computed from the MTG official hotspot product of LSA; the probability density function of the $175$~m resolution ROS of FCI-FireDyn; the Burnt Area (BA) time series of FCI-FireDyn compared to available data from EFFIS, Copernicus EMS and the PT-FireSprd data set; and the Fire Growth Rate (FGR) of FCI-FireDyn. The red strips on the top panel reports time period longer than 20-min with no hotspot detection. Note The EFFIS final BA was estimated $13$ days after the end of the initial propagation.}
\label{fig:aveiroFBM}
\end{figure}
Figure \ref{fig:aveiroFBM} shows the time series of: FRP computed from the sum of all hotspot detected during the duration of the fire from the FCI product\cite{xu2010} of LSA-SAF; the probability density function of the ROS distribution; BA; and FGR. Also reported are the BA estimation from manual delineation of (i) the Copernicus Emergency Monitoring Service (EMS)\cite{EMSR760} that was activated on the 16th at $17$:$35$ UTC, (ii) the PT-FireSprd dataset \cite{Benali2023} that provide sub-daily manual perimeter delineation based on the combination of FCI, polar orbital satellite and airborne observations, (iii) and the EFFIS burn area computed with a semi-automatic algorithm based on visible observation of MODIS and Sentinel-2 data. Since PT-FireSprd provides data at a higher-than-daily frequency, the Fire Growth Rate (FGR) was also computed to enable comparison with the FCI-FireDyn retrieval. 
EFFIS, EMS and PT-FireSprd are using polar orbital visible product (MODIS or Sentinel-2) with far greater resolution than FCI/MTG-I. A lot of details are present within the burnt area, that mask out infrastructure or vegetation that after visual inspection did not burn. To compare those burn area with our approach we apply a morphology closing filter on the delineated burnt area to mimic an equivalent $1.5$~km resolution sensor, hereafter named coarse filter. 
In Figure \ref{fig:aveiroFBM}, BA and FGR time series show that FCI-FireDyn captures the total extent with a slight overestimation ($27,607$ ~ha, between $\approx 2$ to $8\%$ higher than the various coarse estimates from the reference values) at the end time of the propagation (18th at $04$:$00$ UTC). It also pick up the change in growth rate at the time of the outburst ($16$th $06$:$00$ UTC) and during the southern expansion after $00$:$00$ on the $17$th.

\subsection*{Varnavas Fire}
The wildfire that ignited in the region of Varnavas (northeast Attica, Greece) on the 11 August 2024 at 12:00 UTC \cite{EMS746} was driven by extremely hot and dry weather plus strong winds, and spread rapidly toward populated areas covering an area between $10,413$~ha \cite{EMS746} and $10,971$~ha\cite{EFFIS2007}, depending of the source. It was controlled on the 13$^{th}$ of August. 
\begin{figure}[!ht]
\centering
\includegraphics[width=\linewidth]{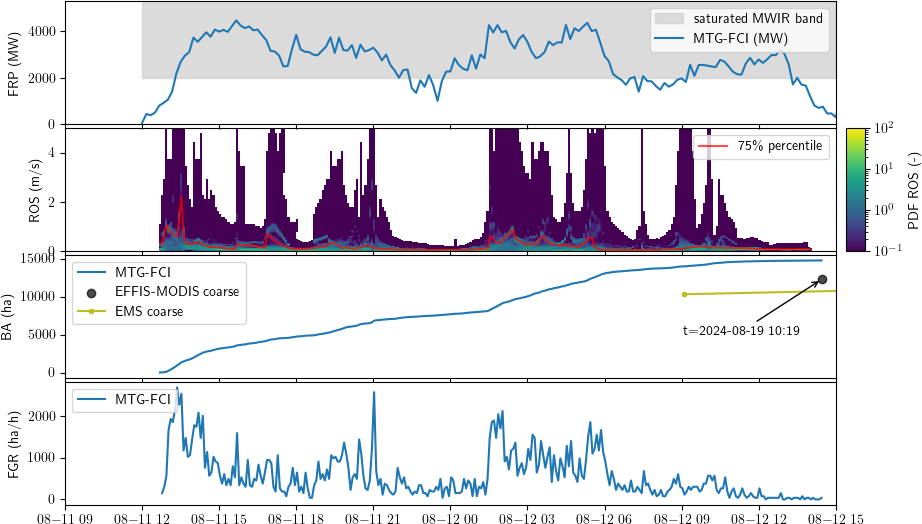}
\caption{Time series of fire-scale fire behavior metrics for the fire that occurred near Varnavas in Greece between the $11^{\text{th}}$ and the $12^{\text{th}}$ of August 2024. From top to bottom panels: the FRP time series computed from the MTG official hotspot product of LSA; the probability density function of the $175$~m resolution ROS of FCI-FireDyn; the Burnt Area (BA) time series of FCI-FireDyn compared to available data from EFFIS, Copernicus EMS; and the Fire Growth Rate (FGR) of FCI-FireDyn. The grey strip on the top panel reports that the MWIR band was saturating at this time for fire temperature and that FRP is underestimated. Note The EFFIS final BA was estimated $7$ days after the end of the initial propagation. }
\label{fig:varnavasFBM}
\end{figure}
At that time FCI was still in commissioning phase, and the MWIR band was saturating when monitoring fire temperature. The FRP time series for the varnavas fire is therefore difficult to interpret. FCI-FireDyn that is using MWIR threshold of $340$ and $360$~K does not show effect from this saturation problem. The final estimated burnt area reads $14,749$~ha, which gives an error between $20$ and $21\%$ higher than the coarse-filtered estimate of the available  EFFIS or EMS burnt area. The ROS distribution computed with FCI-FireDyn shows a spread activity more distributed along the duration of the fire when compared to the Aveiro fire that was showing rather isolated intense peak of ROS. FGR shows also lower maximum values for the Varnavas fire with only several peak above 2000 ha/h.

\subsection*{Ribaute Fire}
\begin{figure}[ht]
\centering
\includegraphics[width=0.5\linewidth]{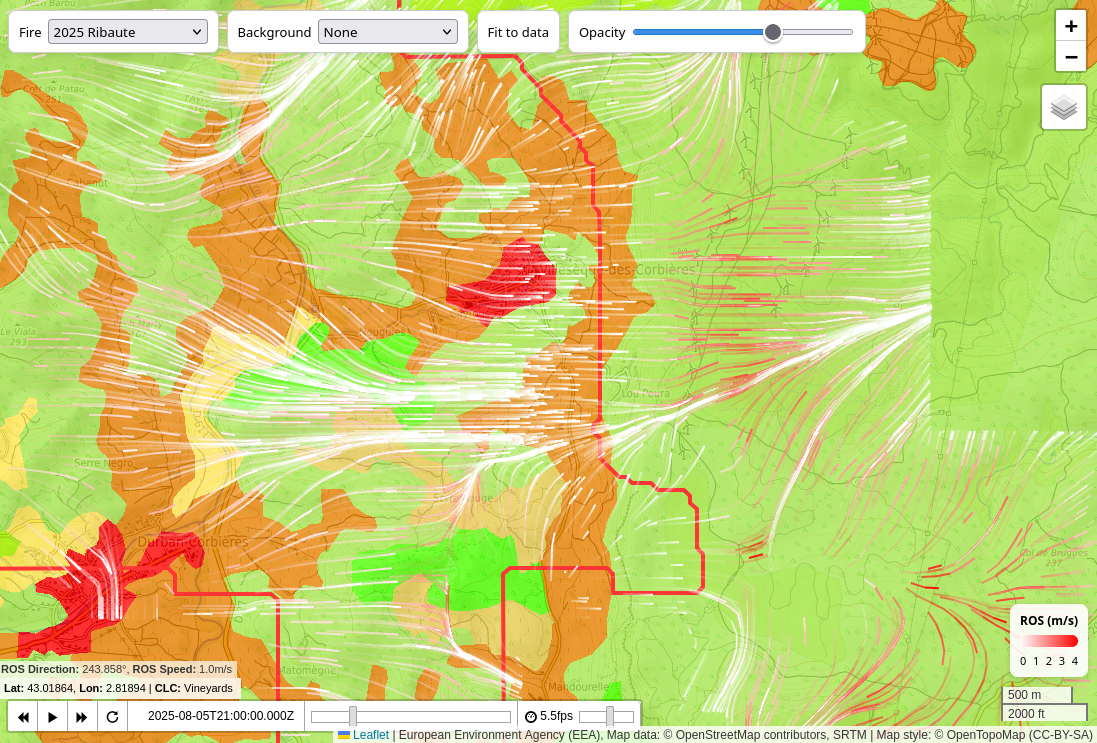}
\caption{View of the fire perimeter (red line) and ROS (color coded on fire path from white to red) for the Ribaute fire at the time of the large Fire Growth Rate (FGR) peak at 21:00 UTC on the $05^{\text{th}}$ of August 2025. The background shows the Corine Land Cover map (red: urban environment, green: Mediterranean maquis, green fluo: mixed forest, brown: vineyards.) This view is an example of the capability of the visualization platform presented in the additional material section.}
\label{fig:ribaute_FGVR6000}
\end{figure}

The fire that occurred on the 5$^{th}$ of August 2025 in the region of Les Corbi\`{e}res in France ended up being the second largest monitored fire in France. The ignition took place near the village of Ribaute at 14:20 UTC, and push by a strong northwesterly wind, it burned $11,391$~ha \cite{EFFIS2007} until 8:00 the next day.
\begin{figure}[!ht]
\centering
\includegraphics[width=\linewidth]{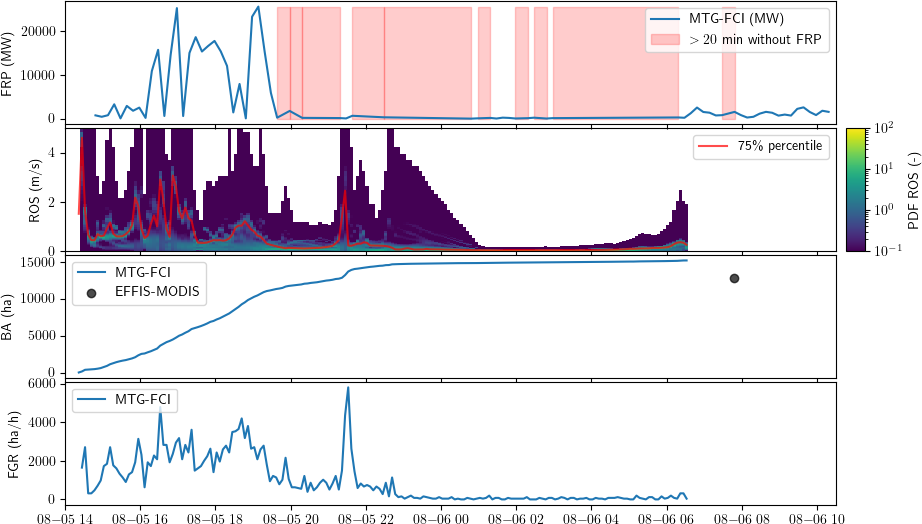}
\caption{Time series of fire-scale fire behavior metrics for the fire that occurred near Ribaute in France between the $5^{\text{th}}$ and the $8^{\text{th}}$ of August 2025. From top to bottom panels: the FRP time series computed from the MTG official hotspot product of LSA; the probability density function of the $175$~m resolution ROS of FCI-FireDyn; the Burnt Area (BA) time series of FCI-FireDyn compared to available data from EFFIS; and the Fire Growth Rate (FGR) of FCI-FireDyn. The red strips on the top panel reports time period longer than 20-min with no hotspot detection. }
\label{fig:ribauteFBM}
\end{figure}
The early build up of the fire during the first two hours shows very high ROS with a 75\% percentile of the ROS close to 1m/s, and a peak within the 20 first min at 4m/s. This very high value as discussed in the next section is probably due to spotting activity that was reported to be very intense at the start of the fire.
Unlike in the Varnavas fire, that also showed a high ROS at the start (75\% ROS $\approx 1m/s$) but followed by relative steadier propagation, in the case of Ribaute, the fire continue to intensify in the next 3h from 16:00 to 19:00 UTC. The 75\% percentile of ROS was then constantly above 1m/s and the FRP show very high variation with peak above 20,000MW. FGR shows during that time a peak at 4000~ha/h. 
The activity slowed down then until 21:00 UTC before a very large increase in FGR at nearly 6000/ha that came with an increase of ROS, but without noticeable change in FRP. This large gain in activity occurred after the head of the fire exited a vineyard area where it was slow down (see Fig \ref{fig:ribaute_FGVR6000}).  

\section*{Discussion and Conclusion}
This manuscript presents FCI-FireDyn, a new algorithm designed to extract fire behavior descriptors at fire front scale such as BA and ROS from the FCI sensor operated onboard MTG-I. Its application to three major fire events in previous section illustrates the algorithm’s capacity to derive fire behavior metrics, including the final burned area from EFFIS or EMS, and the Fire Growth Rate (FGR) derived from sub-daily fire front perimeters provided by the PT-FireSprd dataset \cite{Benali2023}.

\begin{figure}[!ht]
\centering
\includegraphics[width=0.9\linewidth]{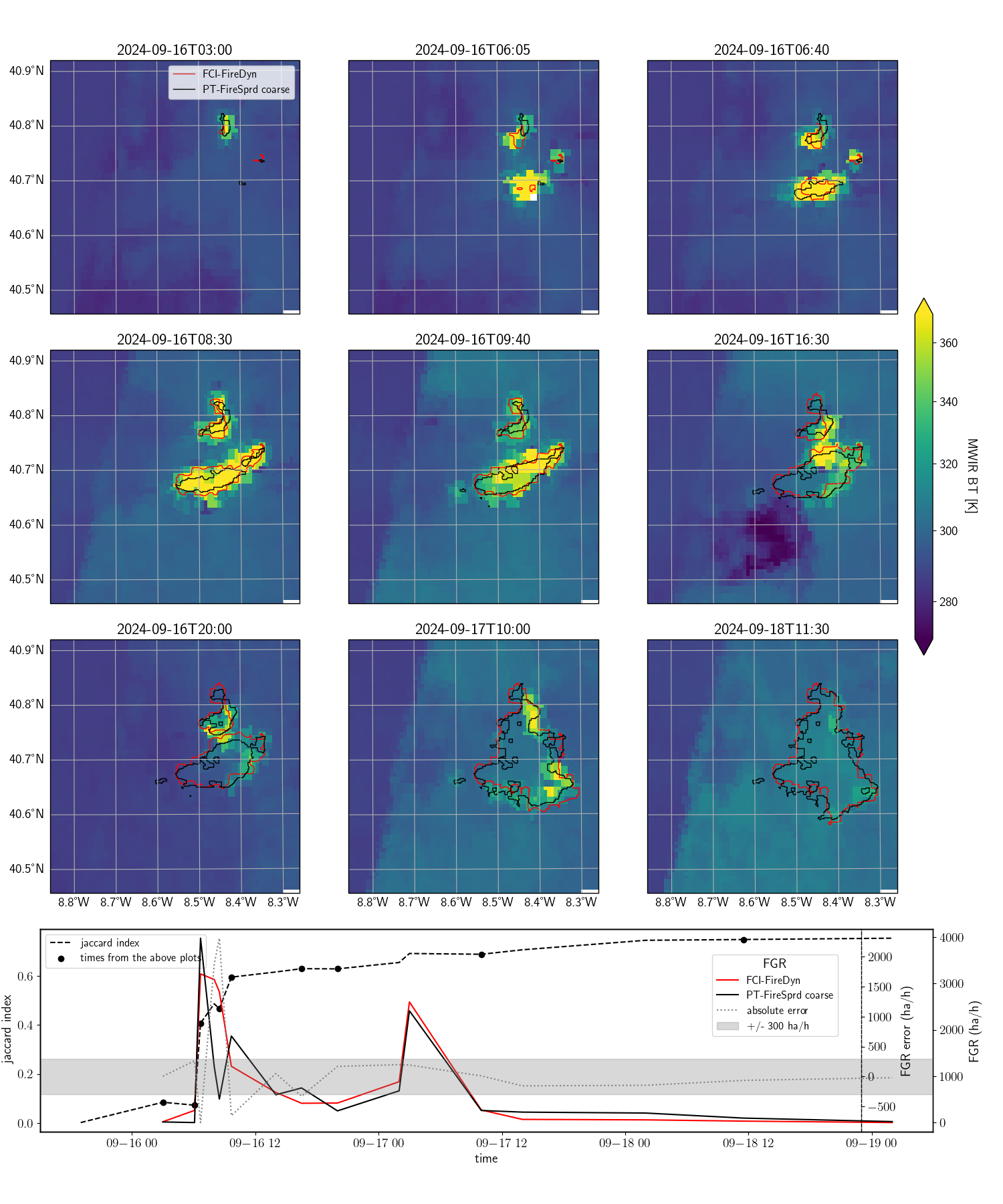}
\put(-450,520){(a)}
\put(-450,120){(b)}
\caption{Panel a shows a comparison of the FCI-FireDyn (red polygon) and the PT-FireSprd (black polygon) fire front perimeters evolution for the Aveiro fire. The background shows the Middle Wave Infra Red images from the FCI sensor. Panel b shows the time evolution of the jaccard index that measure the match between the polygons. and the error in the FGR when calculated on the same selected time.}
\label{fig:aveiro-BA-Evolution}
\end{figure}

The sub-daily front-scale perimeter from PT-FireSprd can also be directly compared with FCI-FireDyn outputs. Figure~\ref{fig:aveiro-BA-Evolution}.a shows the evolution of the fire front perimeter for the Aveiro fire derived from both PT-FireSprd and FCI-FireDyn. PT-FireSprd perimeters are generated through manual interpretation combining FCI imagery and high-resolution airborne data. The coarse filter described earlier is applied to the perimeter of PT-FireSprd.
While the time evolution of Fig \ref{fig:aveiro-BA-Evolution}.a shows the overall good agreement of the BA extracted from the two approaches, Fig \ref{fig:aveiro-BA-Evolution}.b shows a qualitative comparison made possible only for Aveiro due to the availability of sub-daily perimeters.
Fig \ref{fig:aveiro-BA-Evolution}.b is showing time series of the jaccard index between the two fire perimeters and the absolute error between the FGR calculated on the same time sequence. The jaccard index that varies between 0 and 1, is a measure of similariry between perimeters.
Before the 06:40 UTC on the 16-09, the jaccard index is very low ($<0.2$). Then the fire get bigger ($\text{BA}>2000$~ha) and there is an increase of the jaccard index that remain above $0.6$ after 09:40 UTC. Meanwhile the FGR comparison shows an absolute error within $\pm 300$~ha/h during the all fire except during the episode of the outburst after 06:00 UCP on the 16-09. The two sharp increases of the FGR are however at the right time, only the evolution of the outburst episode from 06:00 to 09:40~UTC shows large difference.
At 09:40~UTC the PT-FireSprd perimeter includes airborne data interpretation. It aligns closely with FCI-FireDyn and the subsequent dynamics of the fronts shows good agreement, in particular at the time of the southern expansion after 00:00 UTC on the 17-09. 
It shows that FCI-FireDyn does not perform well before the fire exhibit a certain propagation run (the early stage of the aveiro fire is poorly captured). However, once the fire is running, the similarity of the perimeter is rather good considering the difference in the spatial resolution of the input data and the timing of the activity changes agree well. Further test on fire from he PT-FireSprd would be considered in the future.

Among the various fire front dynamic metrics—FSR, CROS, and FROS—only the latter can be directly associated with front-scale processes, whereas the others reflect rather fire-scale properties. A quantitative comparison between ROS estimated by FCI-FireDyn and FROS derived from PT-FireSprd \cite{Benali2023} or Duane et al. (2024) \cite{Duane2024} remains however challenging. While both represent fire spread at front scale, they capture different aspects of propagation. The fire paths (or fire propagation axis) retrieved by FCI-FireDyn are based on neighboring arrival times from 10-minute, 1~km resolution MWIR observations combined with spatio-temporal interpolation. In contrast, FROS-based approaches \cite{Benali2023,Duane2024} average over longer time intervals and larger spatial scales, relying primarily on geometric analysis rather than explicit fire activity metrics.
\begin{figure}[ht]
\centering
\includegraphics[width=1.\linewidth]{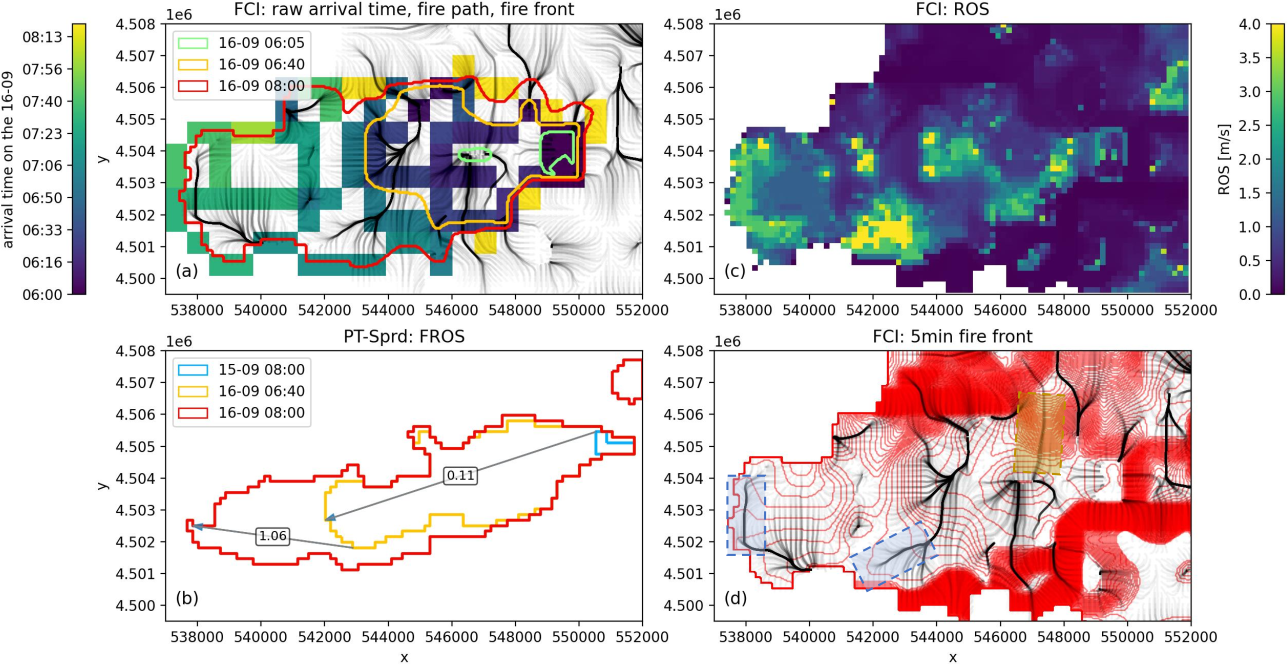}
\caption{Comparison between the Rate of Spread (ROS) derived from FCI-FireDyn and the Forward Rate of Spread (FROS) obtained from PT-FireSprd in the context of the fire outburst that occurred at 06:00~UTC on the $16^{\text{th}}$ of September~2024 during the Aveiro fire. Panel~(a) shows the low-resolution sparse initial arrival time map together with high-resolution fire front selected along the duration of the outburst event (colored polygon) and high resolution fire path (black line). Panel~(b) presents available perimeter from PT-FireSprd during the outburst with the coarse spatial filter applied and the associated FROS. . Panels~(c) and (d) display outputs from FCI-FireDyn, namely the high-resolution ROS field~(c), and the corresponding fire front and fire path~(d). In Panel (d), the fire-front interval is displayed at 5 minutes for visualization clarity. The shaded blue and yellow boxes mark the areas of head or finger fire activity and of front merging, respectively—both identified by the convergence of the high-resolution fire paths.}
\label{fig:compa-fci-ptsprd}
\end{figure}
Figure~\ref{fig:compa-fci-ptsprd} illustrates these differences for the Aveiro fire during the outburst after 06:00~UTC on 16 September. Discrepancies of approximately 2~km in intermediate perimeters and slight shifts in ignition time that are linked to the low value of the jaccard index in Fig \ref{fig:aveiro-BA-Evolution}.b are mostly attributed to uncertainties in arrival time estimation from the 10-minute MWIR sequence. In addition to the inherent detection delay of FCI compared to airborne data, that are due to its resolution, FCI-FireDyn introduces an additional delay by defining a pixel’s arrival time only when the MWIR brightness temperature (BT) remains above a threshold for at least 30~minutes (three consecutive frames). This conservative approach minimizes false detections caused by FCI’s slight image jitter. Consequently, FCI-FireDyn may introduce delay in rapid fire expansion phases (like this outburst); however, this should have limited impact on average ROS and BA estimation when ROS remains approximately stable over 30-minute periods.

The core strength of FCI-FireDyn lies in its spatio-temporal interpolation logic, which constructs arrival time maps, critical in the downscaling. This approach allows retrieval of detailed fire dynamics, delineating main propagation axes where fire paths converge—either during front merging (green box in Fig.~\ref{fig:compa-fci-ptsprd}c) or during head and finger front expansion (blue boxes in Fig.~\ref{fig:compa-fci-ptsprd}). 

FCI-FireDyn estimated ROS values obtained for the three test fires fall within the expected ranges of fire behavior categories reported by \cite{Chuvieco2024}.
\begin{figure}[ht]
\centering
\includegraphics[width=0.5\linewidth]{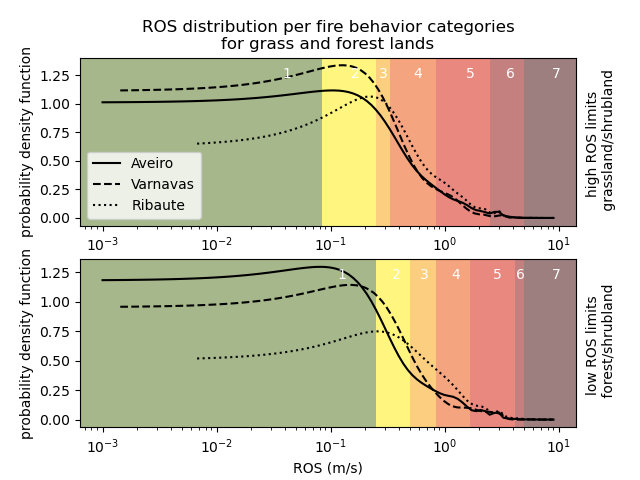}
\caption{Probability Density Function (PDF) of the Rate of Spread (ROS) for the three fire scenarios (Aveiro, Varnavas, Ribaute) and the two dominant vegetation types (grassland and forest). The shaded background represents the fire behavior categories defined in \cite{Chuvieco2024}, illustrating the distribution of ROS values across varying fire intensity regimes.}
\label{fig:pdf_ros_fires}
\end{figure}
Figure~\ref{fig:pdf_ros_fires} shows the ROS probability density functions (PDF), overlaid over the fire category classification from \cite{Chuvieco2024}, for the three fire events split per grass and forest dominated land cover. While all events exhibit a mix of categories, the Ribaute fire shows a higher prevalence within the intense classes (categories~4–5) for both grass and forest lands. This is a relative comparison between the those three large scale event, all reporting extreme fire scenario. It is emphasizing the exceptional behavior of the Ribaute fire. 

When interpreting fire behavior, the peaks in FGR generally coincide with increases in FRP, although the FRP peaks tend to be broader in time (Fig \ref{fig:aveiroFBM}). This indicates that the emitted energy is not solely associated with the advancing fire front, smoldering and cooling zones contribute to the overall radiative signal. Notably, each major increase in FRP is preceded by a corresponding rise in the ROS distribution, as indicated by increases in the 75th percentile of the ROS distribution (Fig \ref{fig:aveiroFBM}). This suggests that enhanced energy release is linked to the expansion of the actively burning area. Increase of FRP and therefore emission comes with spread, but sustain emission can remain after phase of propagation.    


The derived FCI-FireDyn metrics have potential applications for comparison with predictive modeling of fire spread and in improving emission in atmospheric transport models used for smoke dispersion estimation. The perimeter evolution can be used to refine existing algorithms of fire emissions \cite{Turquety2020} by providing spatially explicit, 175~m high resolution cell-to-cell fire spread estimates based on observational evidence.

The FCI-FireDyn workflow consists of four main steps: 
(i) computation of the first arrival time map, 
(ii) estimation of an initial low resolution ROS vector field, 
(iii) downscaling of the arrival time map using fire path created from the ROS vector field, 
and (iv) generation of the final ROS and fire front perimeters dynamics
(see Methods section). 
Currently, the first step relies solely on the MWIR band, as the FCI active fire (hotspot) product was not available at the time of analysis. Future developments will focus on reducing the latency of this step. In particular, the threshold values of the arrival time calculation will be better justified, for example by including atmospheric correction to the MWIR signal or by estimating threshold values per land cover type. 
The upcoming MTG-I 2 imager, providing 2.5-minute temporal resolution over Europe, will also further enhance timing precision.
Parallax distortions caused by plume geometry may influence MWIR signatures, potentially leading to BA and ROS overestimation in some scenarios. Higher temporal sampling from MTG-I 2 and inclusion of hotspot information could help discriminate active fire fronts from plume dynamics. 


In several cases—most clearly during the Aveiro fire—rapid increases in ROS were observed to precede corresponding rises in FRP. This behavior indicates that FCI-FireDyn could, in principle, contribute to the early detection of changes in fire behavior when integrated into NRT workflows. However, accurate ROS estimation depends on the interpolation of the arrival-time map and therefore requires at least one subsequent observation that marks the end of the current activity phase. For operational use, for example, sub-pixel front displacements would need to be resolved to prevent misclassifying slowly advancing fronts as inactive. At present, FCI-FireDyn is not yet suitable for NRT implementation and will require further development to enhance its ability to interpret fire-front dynamics in real time.

Overall, this study provides the first demonstration of MTG-FCI’s capability to retrieve high-resolution fire dynamics metrics—ROS, FGR, and BA/perimeter—at fire front scale. Despite inherent limitations related to spatial resolution and temporal latency, the results show that FCI-FireDyn captures essential features of large fire event evolution, with strong potential for both scientific and operational applications.

\section*{Methods}
This section outlines the algorithms underlying FCI-FireDyn. It begins by describing the spatio-temporal interpolation approach used to derive a ROS vector field from a sparse arrival time map. This interpolation method is applied twice within the four main steps that form the core of FCI-FireDyn, which are introduced sequentially:
(i) computation of the first arrival time map, 
(ii) estimation of an initial low resolution ROS vector field, 
(iii) downscaling of the arrival time map using fire path created from the ROS vector field, 
and (iv) generation of the final ROS and fire front perimeters dynamics.

\subsection*{ROS computation: arrival time map to ROS}
A new approach for ROS computation is proposed here. The classical approach\cite{Paugam2013,Johnston2018,Stow2019} seeks to estimate distance of propagation between two fire front location.  Here we work on time estimation rather than  distance using interpolation of the sparse arrival time map.
For this purpose, we start with an arrival time map, $t_{\text{arrT}}$ of spatial resolution $dx$ . It is formed from the combination of all the observed local arrival time map, and may have gaps. 
To complete the sparse arrival time map $t_{\text{arrT}}$, a two-step interpolation approach is used. Fire front perimeter $\Gamma^{t}$ are defined from the sparse arrival time map as contour where $t_{\text{arrT}}=t$. 

First, a linear interpolation computes the mean arrival time in points of each gap, formed between two consecutive front perimeters $\Gamma^{t}$ and $\Gamma^{t+1}$,
by considering all possible linear forward propagation scenarios from points belonging to $\Gamma^{t}$ towards all accessible points from $\Gamma^{t+dt}$, and inversely all possible backward propagation scenarios ($\Gamma^{t+dt}$ to $\Gamma^{t}$). The linear propagation must verified three conditions: 
($i$)~a mean ROS smaller than a fixed maximum value (\eg $\text{ROS}^{max}_{\text{lin}}=10$~m\,s$^{-1}$ twice the lower limit of the higher fire behavior category \cite{Chuvieco2024}); 
($ii$)~direct linear connection between the two points; 
and ($iii$)~a distance smaller than one and a half times the smallest dimension of the gap and within one and a half standard deviations to the mean distance of all possible distances in the gap. 

Second, an additional interpolation based on a Radial Basis Function (RBF) with a multiquadric kernel is used to fill any remaining gaps. The local shift between the overall spatial trend of $t_{\text{arrT}}$ and the values in $t_{\text{arrT}}$ surrounding a gap, induce local depressions (“dips”) or elevations (“bumps”) in the interpolated surface, which correspond respectively to secondary ignitions or to the merging of fire fronts.
The experience shows that RBF is usually call to generate secondary ignition that are assigned to explain gap that could not be filled up with the previous linera interpolation.

The normalized gradient of the fully interpolated arrival time map $t_{\text{arrT}}$ is then used to compute a map of local fire front propagation direction defined by $\vec{n} =\nabla t_{\text{arrT}} / \norm{\nabla t_{\text{arrT}}}$. 
The underlying assumption is that the fire propagate normal to the front as done in fire front modeling approach like level set model \cite{Mallet2009} or Lagrangian method \cite{Filippi2013}.
Local ROS is  calculated in this propagation direction by extracting the time from the arrival time map required to propagate over a length equal to the map resolution ($dx$). Finally, the local ROS value is defined as the mean of the forward and backward (calculated using $-\vec{n}$) ROS values as in \cite{Johnston2018}.

This ROS computation methodology has the advantage of not relying on the determination of spread vector computed from only a pair of points located on successive fire front as in \cite{Johnston2018} or \cite{Stow2019}. Here the spread direction is computed using multiple pairs of points, creating smoother contours.

\subsection*{FCI-FireDyn: (i) Computation of the initial Arrival Time Map}

The initial arrival time map, denoted as $t_{\text{arrT}}^{\text{lr}}$, is derived from the low resolution mid-wave infrared (MWIR) band ($3.8,\mu\mathrm{m}$) of FCI resampled at $768$~m (half of the native resolution of FCI over latitude of southern Europe). The estimation relies on a straightforward threshold-based method applied to the pixel brightness temperature time series $T_{\text{MWIR}}^{xy}(t)$ extracted at location $(x,y)$.

A two-step thresholding approach is adopted, using two temperature thresholds $T_{\text{MWIR}}^1$ and $T_{\text{MWIR}}^2$. A pixel is considered to be actively burning if its brightness temperature $T_{\text{MWIR}}^{xy}(t)$ remains continuously above $T_{\text{MWIR}}^1$ for at least 30 minutes. The corresponding arrival time $t_{\text{arrT}}^{\text{lr}}[x,y]$ is then defined as the time of the next local maximum in $T_{\text{MWIR}}^{xy}(t)$ that exceeds the higher threshold $T_{\text{MWIR}}^2$.

In the current version, the brightness temperature series $T_{\text{MWIR}}^{xy}(t)$ are not corrected for atmospheric transmittance. Fixed threshold values of $T_{\text{MWIR}}^1 = 340$~K and $T_{\text{MWIR}}^2 = 360$~K are however used for all three fire scenarios. An atmospheric correction procedure will be implemented in a subsequent version. The values of $T_{\text{MWIR}}^1$ and $T_{\text{MWIR}}^2$ were selected to best adapt to the three fire scenario of this study.

When a cell $(x,y)$ is not assigned with an arrival time, but $T_{\text{MWIR}}^{xy}(t)$ remains above $T_{\text{MWIR}}^1$ for 30 minutes, $t_{\text{arrT}}^{\text{lr}}[x,y]$ is set to undefined. Consequently, $t_{\text{arrT}}^{\text{lr}}$ is a sparse matrix.

\subsection*{FCI-FireDyn: (ii) estimation of an initial low resolution ROS vector field}
The second step consist in using the sparse arrival time map $t_{\text{arrT}}^{\text{lr}}$, to compute a ROS vector field $\text{ROS}^{\text{lr}}$ based on the ROS computation introduced above. 

\subsection*{FCI-FireDyn: (iii) Downscaling arrival time map using fire path created from $\text{ROS}^{\text{lr}}$.}
Streamlines are linearly interpolated from $\text{ROS}^{\text{lr}}$ field vector to generate fire paths using an iterative 2-minute time step. The starting points of these fire paths are taken from a regular 175 m grid covering the same spatial extent as the original data (i.e., 1/4 of the resampled resolution). Fire locations along these paths are then extracted at 1-minute intervals to construct a refined arrival time map at the spatial resolution of $175$~m,  $t_{\text{arrT}}^{\text{hr}}$, that also comes as a sparse matrix. At this stage a water body mask is incorporated into the arrival time computation to exclude part of the fire path that reach non-burnable areas.  

\subsection*{FCI-FireDyn: (iv) Generation of the final ROS and fire front perimeters dynamics}

The same approach as in Step (ii) is applied to \( t_{\text{arrT}}^{\text{hr}} \) to produce a high-resolution fire arrival time map and the corresponding ROS vector field, \( \text{ROS}^{\text{hr}} \).  

The main outputs of FCI-FireDyn are the \( \text{ROS}^{\text{hr}} \) field---visualized within the interactive platform using particle motion to highlight fire path convergence during head fire development or front merging---and the dynamic fire front perimeters, which can be extracted from the interpolated \( t_{\text{arrT}}^{\text{hr}} \). A 2-min interval was selected to compute the fire front. Higher value were not enough precise to capture the rapid fire front change in the location of high ROS values.  

\bibliography{referencesMendeley}



\section*{Acknowledgements (not compulsory)}
This work was partly funded by Grant PID2023-150607OB-I00 funded by MICIU/AEI/27610.13039/501100011033
and the European Union under the Interreg Sudoe Programme (S2/2.4/F0327).


\section*{Additional information}
To enable detailed analysis at the fire front scale, an online visualization platform was developed and is accessible at \url{https://certec.mtg.eebe.upc.edu/}.  
The platform displays for the three fire scenarios:  
\begin{itemize}
    \item reconstructed fire fronts (bold red lines) at 2-minute intervals,  
    \item Rate of Spread (ROS) streamlines, color-coded from white to red according to magnitude,  
    \item Corine Land Cover (CLC) information,  
    \item the first Sentinel-2 true color image acquired after the fire,  
    \item the EFFIS burnt area from a semi-automatic algorithm based on the MODIS/SENTINEL-2 data,
    \item the MWIR band of FCI at 10-minute intervals, resampled at $768$~m,  
    \item the FCI true-color composite imagery at 10-minute intervals also resampled at $768$~m.  
\end{itemize}

To include, in this order: \textbf{Accession codes} (where applicable); \textbf{Competing interests} (mandatory statement). 

The corresponding author is responsible for submitting a \href{http://www.nature.com/srep/policies/index.html#competing}{competing interests statement} on behalf of all authors of the paper. This statement must be included in the submitted article file.

\end{document}